\documentclass[a4paper,11pt]{article}
\usepackage{pos}
\usepackage[utf8]{inputenc}
\usepackage{textgreek}
\usepackage{lineno}
\usepackage{hepnames}
\title{Overview of recent UPC measurements}

\author[a]{Anisa Khatun on behalf of the ALICE Collaboration}

\affiliation[a]{University of Foggia and INFN\\
Foggia, Italy}

\emailAdd{anisa.khatun@cern.ch}
\abstract{
This contribution presents an overview of recent measurements of photon-induced processes in ultra-peripheral collisions (UPCs) performed with the ALICE experiment at the LHC. Results from Run~2 include detailed studies of exclusive vector meson photoproduction in Pb--Pb collisions. Measurements of incoherent J/$\psi$ photoproduction as a function of energy and the Mandelstam-$|t|$ variable, combined with electromagnetic dissociation (EMD) classification, show a suppression of the energy evolution at large $|t|$, favouring saturation-based descriptions of the gluon structure. Complementary measurements of proton emission in Pb--Pb UPCs constrain nuclear breakup mechanisms and provide a handle on collision geometry through EMD tagging. Coherent $\rho^{0}$ photoproduction exhibits impact-parameter-dependent azimuthal anisotropy consistent with quantum interference effects, while the first polarisation measurement of coherently photoproduced J/$\psi$ confirms $s$-channel helicity conservation. Additional measurements of exclusive multi-hadron photoproduction, including four-pion and charged kaon pair final states, probe resonance contributions and light vector meson couplings to photons and nuclear targets.

Run~3 data enable the study of inclusive photonuclear interactions. Measurements of inclusive open charm photoproduction constrain the gluon content of nuclei in the perturbative QCD regime. Studies of identified hadron production and baryon-to-meson ratios in photonuclear collisions provide new information on particle production mechanisms and possible collective effects in photon-induced systems. Photon--photon interactions are also studied through performance and sensitivity analyses for the measurement of the anomalous magnetic moment of the tau lepton, together with projections for future measurements with ALICE~3.

}

\FullConference{
}

\begin{document}
\maketitle

\section{Introduction}
Ultra-peripheral collisions (UPCs) of heavy ions at the LHC provide a clean and controlled environment to study photon-induced interactions at unprecedented energies. In these collisions, the impact parameter exceeds the sum of the nuclear radii, strongly suppressing hadronic interactions while allowing electromagnetic interactions to dominate. The strong electromagnetic fields generated by relativistic heavy ions can be described as fluxes of quasi-real photons, enabling both photon--photon and photon--nucleus interactions.

UPCs have emerged as a powerful tool to address a wide range of physics topics, including the structure of nucleons and nuclei, the behavior of gluon distributions at small Bjorken-$x$, soft and hard QCD dynamics, and electroweak processes. The large photon flux, scaling with the square of the nuclear charge, allows access to rare processes with measurable cross sections.

At ALICE, UPCs are characterized by a small number of final-state particles in an otherwise empty detector, often accompanied by electromagnetic dissociation (EMD) of one or both nuclei measured using Zero Degree Calorimeters (ZDC)~\cite{ALICERun2}. The detection of forward neutrons and protons provides experimental access to the impact parameter of the collision, enabling geometry-dependent studies. The transverse momentum of the produced system naturally distinguishes coherent photon--nucleus interactions, where the nucleus remains intact, from incoherent and dissociative processes probing nucleonic and subnucleonic fluctuations.

Over the past decade, ALICE has established a comprehensive UPC physics program~\cite{Alice2}. In Run~3, the scope of UPC studies has expanded significantly, allowing not only precision measurements of exclusive processes but also the first systematic investigations of inclusive photonuclear interactions. In particular, UPCs at the LHC provide access to Bjorken-$x$ values far below those reachable in fixed-target and lepton--hadron experiments. 



\section{UPCs as a Probe of Nuclear Structure and QCD}

\begin{figure}[b!]
     \includegraphics[width=0.56\textwidth]{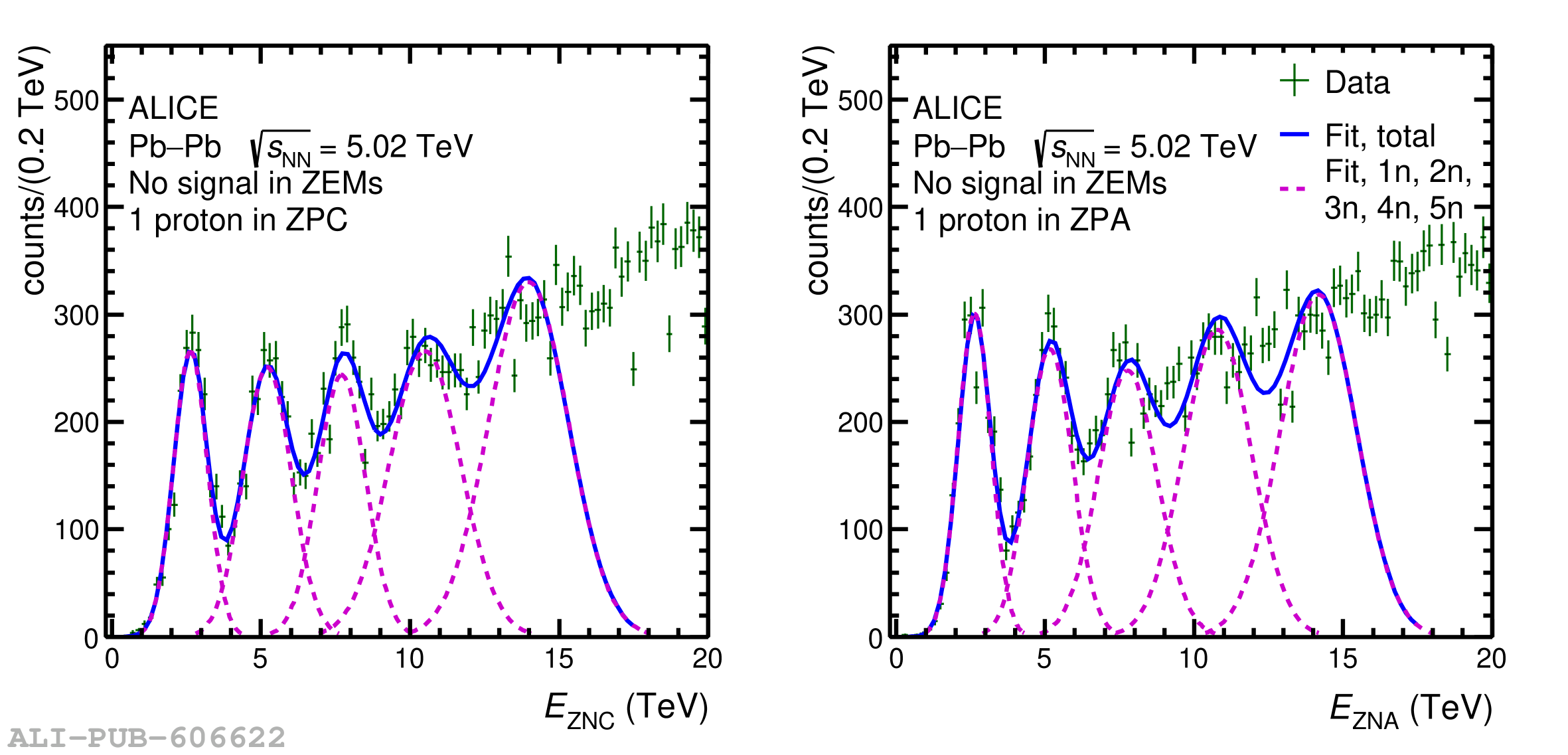}
    \includegraphics[width=0.43\textwidth]{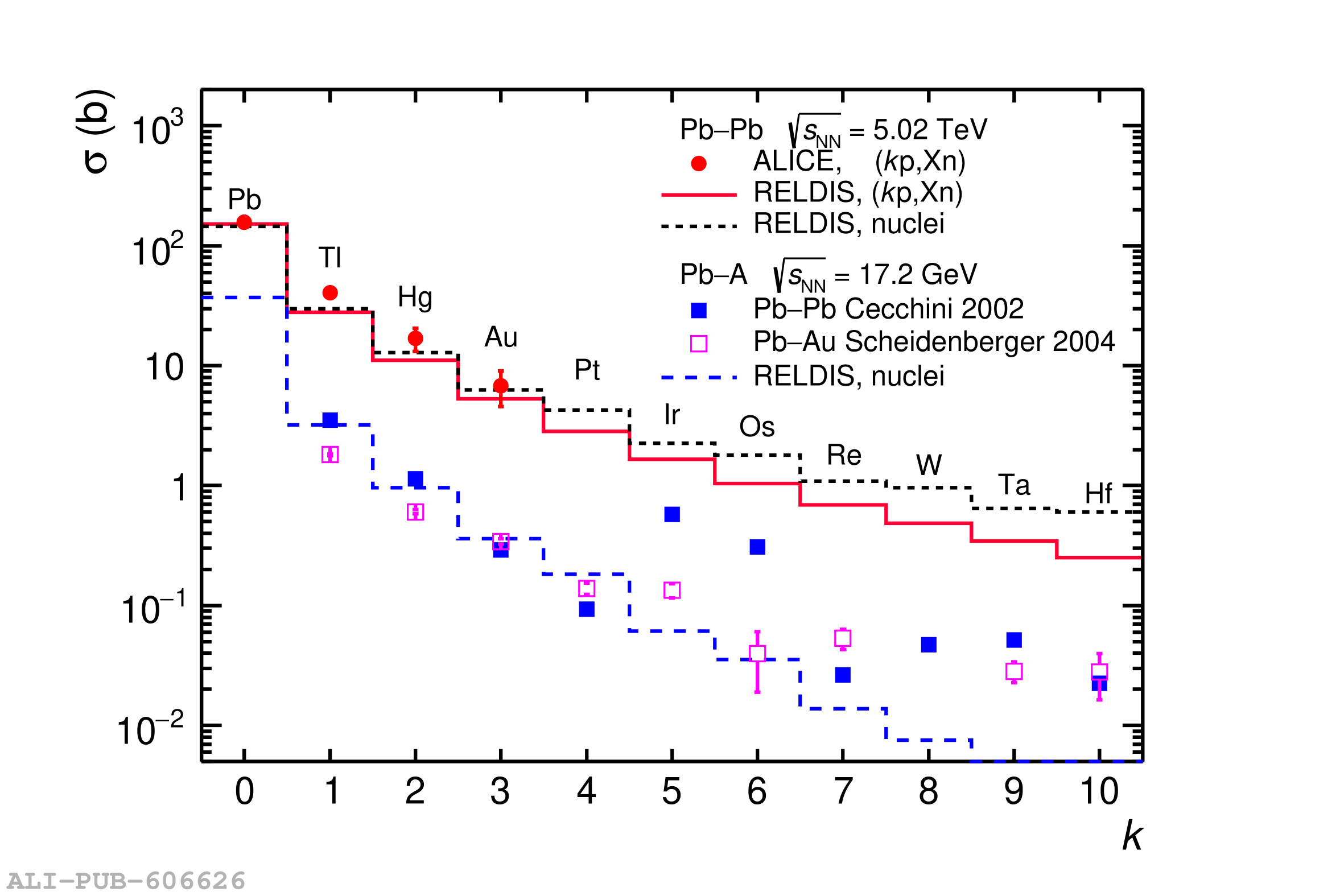}
   
    \caption{(a) Energy distributions for single-proton emission accompanied by neutrons measured in ZNC (left) and ZNA (right) in EMD events. (b) Isotope production (Pb, Tl, Hg, Au) in nuclear breakup in Pb--Pb UPCs~\cite{protonemission}.}

    \label{fig:goldproduction}
\end{figure}

 ALICE has performed the first measurement of proton emission accompanied by neutrons in EMD of Pb--Pb UPCs, as shown in the left and middle panels of Fig.~\ref{fig:goldproduction}, in addition to neutron emission~\cite{neutronemission, neutronemission2, protonemission}. 
 The energy distributions measured in the neutron calorimeters (ZNC/ZNA) of the ZDC system, for events tagged with a single proton in the proton calorimeters (ZPC/ZPA) and with no signal in the electromagnetic ZEM detectors ($4.8<\eta<5.7$), demonstrate clear sensitivity to fragments emitted at very forward rapidities ($|\eta|>8.8$).
 Comparisons with the RELDIS model show a reasonable description of the 0 and 3 proton channels, while the yields for 1 and 2 proton emission are underestimated by about 20\%. Mixed channels with 1, 2, 3 neutrons besides 1 proton emission are overestimated, indicating limitations in current descriptions of proton evaporation and nuclear breakup~\cite{protonemission}. Proton emission in EMD provides direct sensitivity to isotope production in the nuclear breakup of Pb nuclei, including Tl, Hg, and Au. In particular, ALICE has observed the production of gold nuclei via the emission of three protons from $^{208}$Pb, with a measured cross section of $\sigma_{\rm 3p} = 6.8 \pm 2.2~\mathrm{b}$, comparable to the total inelastic Pb--Pb hadronic cross section at the LHC~\cite{protonemission}. Beyond its striking transmutation aspect, this result establishes proton-tagged EMD as a clean and powerful handle on collision geometry, significantly enhancing photonuclear studies based on EMD classes.

One of the central motivations for UPC studies is the investigation of nuclear structure and QCD dynamics at high energies. Exclusive vector meson photoproduction provides sensitivity to the gluon content of the target nucleus, since in leading-order descriptions the production amplitude is related to the gluon distribution of the target~\cite{EPhJC822022413}. More recent next-to-leading order (NLO) calculations indicate sizable corrections and a more complex dependence on the gluon distribution~\cite{JEskola}. These measurements therefore provide constraints on nuclear parton distribution functions and on phenomena such as gluon shadowing and saturation~\cite{EPCJ752015580}.

ALICE has measured coherent and incoherent photoproduction of heavy quarkonia, notably J/$\psi$, in Pb--Pb UPCs at both midrapidity and forward rapidity~\cite{PLB798, midrapjpsi, JHEP10, firstCohmant, firstInCohmant}.  These measurements probe Bjorken-$x$ values down to approximately $10^{-5}$, a region where nonlinear QCD effects may become relevant. The separation of coherent and incoherent contributions, based on transverse momentum ($p_{\rm T}$) distributions, allows the study of average nuclear gluon densities as well as their spatial fluctuations~\cite{EPJC74, firstCohmant, firstInCohmant}. More details on this topic can be found here~\cite{low-xAK}.

\begin{figure}
    \includegraphics[width=0.98\textwidth]{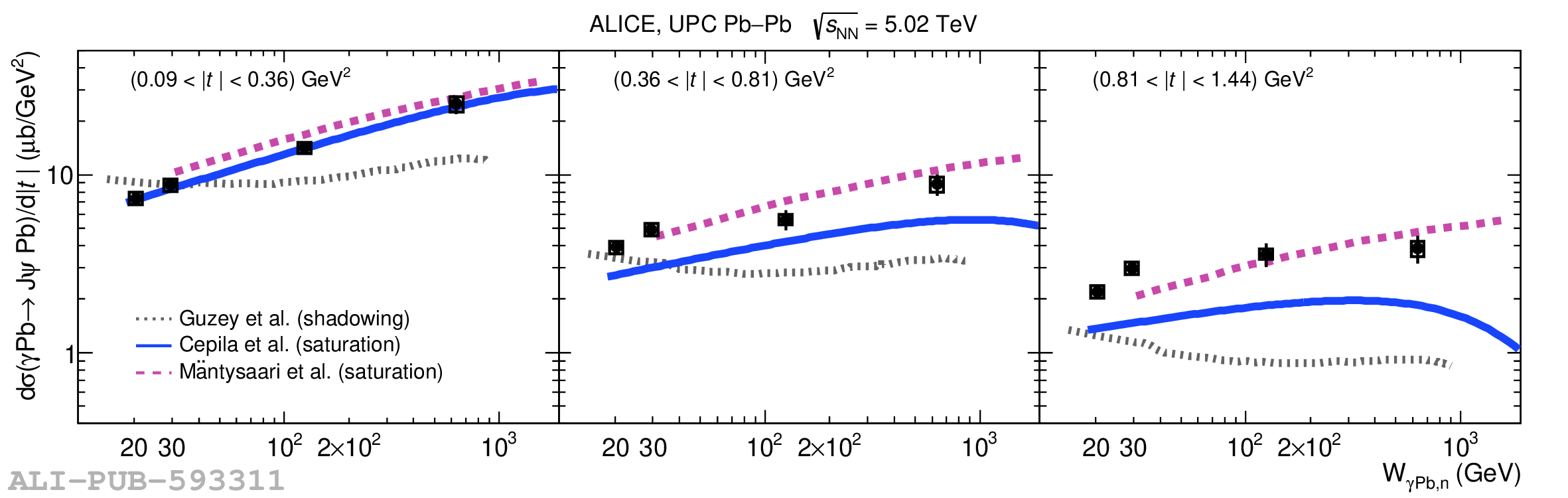}
    \caption{Energy dependence of the incoherent J/$\psi$ photonuclear cross section off Pb for three $|t|$ intervals, compared with shadowing- and saturation-based model predictions~\cite{gluonsaturation}.}
    \label{fig:sup}
\end{figure}

Building on the geometric sensitivity provided by EMD tagging, ALICE has recently performed multi-differential measurements using electromagnetic dissociation classes~\cite{PRL89, firstCohmant, firstInCohmant}. By combining the quadri-momentum transfer of the reaction, $|t|$, with neutron emission tagging, sensitivity to the impact-parameter dependence of photonuclear interactions is achieved. As shown in Fig.~\ref{fig:sup}, recent ALICE measurements show evidence for a suppression of the growth of incoherent J/$\psi$ production with energy at large $|t|$, favoring saturation-based models over descriptions based solely on nuclear shadowing~\cite{gluonsaturation}. These results highlight the unique role of UPCs in accessing the spatial structure of gluon fields at nucleonic and subnucleonic scales.

\section{Light Vector Mesons and Spin}
The photoproduction of light vector mesons provides complementary insight into soft QCD dynamics, quantum interference effects, and spin phenomena. ALICE has performed detailed studies of coherent $\rho^{0}$ and di-kaon meson photoproduction in heavy-ion UPCs, exploiting the excellent low-transverse-momentum acceptance of the detector~\cite{rhoxexe,rhopbpb,dikaon}.

A particularly striking result is the first measurement of impact-parameter-dependent azimuthal anisotropy in coherent $\rho^{0}$ photoproduction ~\cite{azimuthalanisotropy}. Using electromagnetic dissociation classes as a proxy for collision geometry, ALICE observed a strong $\cos(2\varphi)$ modulation that increases by nearly an order of magnitude from large to small impact parameters. This behavior is interpreted as a quantum interference effect arising from the indistinguishability of photon emission from either nucleus and demonstrates sensitivity to femtoscopic-scale geometry in UPCs~\cite{azimuthalanisotropy}. 

In addition, ALICE has reported the first polarization measurement of coherently photoproduced J/$\psi$ mesons in Pb--Pb UPCs at $\sqrt{s_{\rm NN}} = 5.02$~TeV, measured via the dimuon decay channel at forward rapidity~\cite{jpsipolarisation}. The polarization parameters $\lambda_{\theta}$, $\lambda_{\varphi}$, and $\lambda_{\theta\varphi}$ were extracted in the helicity frame and are found to be consistent with transverse polarization, in agreement with $s$-channel helicity conservation. The results are compatible with previous measurements in ep collisions at lower energies, providing the first evidence that coherent J/$\psi$ photoproduction off Pb nuclei follows the same helicity dynamics as observed in lepton--hadron interactions. These measurements demonstrate that UPCs probe not only the density and geometry of gluons, but also the fundamental production mechanisms and spin structure of vector meson photoproduction.

ALICE has also performed detailed studies of low-mass vector meson photoproduction. Differential measurements of coherent charged-kaon-pair photoproduction in Pb--Pb UPCs provide access to the coupling of the $\phi(1020)$ meson to photons and nuclear targets~\cite{dikaon}. In addition, ALICE has measured coherent photoproduction of four-pion final states, where the data are well described by the interference of the $\rho(1450)$ and $\rho(1700)$ resonances~\cite{4pion}. Owing to its intermediate mass, both channels probe gluon dynamics in a regime harder than $\rho^{0}$ photoproduction while remaining complementary to quarkonium measurements. Together, these measurements establish UPCs as a versatile laboratory for studying vector meson production across a broad range of mass scales and QCD regimes.

\section{The ALICE Experiment and UPCs in Run~3}
The ALICE detector is uniquely suited for UPC studies due to its excellent tracking, particle identification, and forward veto capabilities. In Run~3, ALICE underwent a major upgrade featuring continuous readout, an improved Inner Tracking System, and enhanced forward detector coverage, significantly extending the UPC physics reach of the experiment~\cite{trigupdate, alice3,yellowrep}.

For UPC analyses, the upgraded Forward Interaction Trigger (FIT) and ZDC provide increased flexibility in event selection. In contrast to Run~2, Run~3 allows asymmetric vetoing of forward detector activity, enabling the selection of exclusive, semi-inclusive, and inclusive UPC topologies~\cite{fitupgrade, Alic3upc}. This capability opens a new window for studying inelastic photonuclear interactions and small-system phenomena, extending the UPC program beyond the predominantly exclusive measurements of Run~2.

\begin{figure}
    \includegraphics[width=0.49\linewidth]{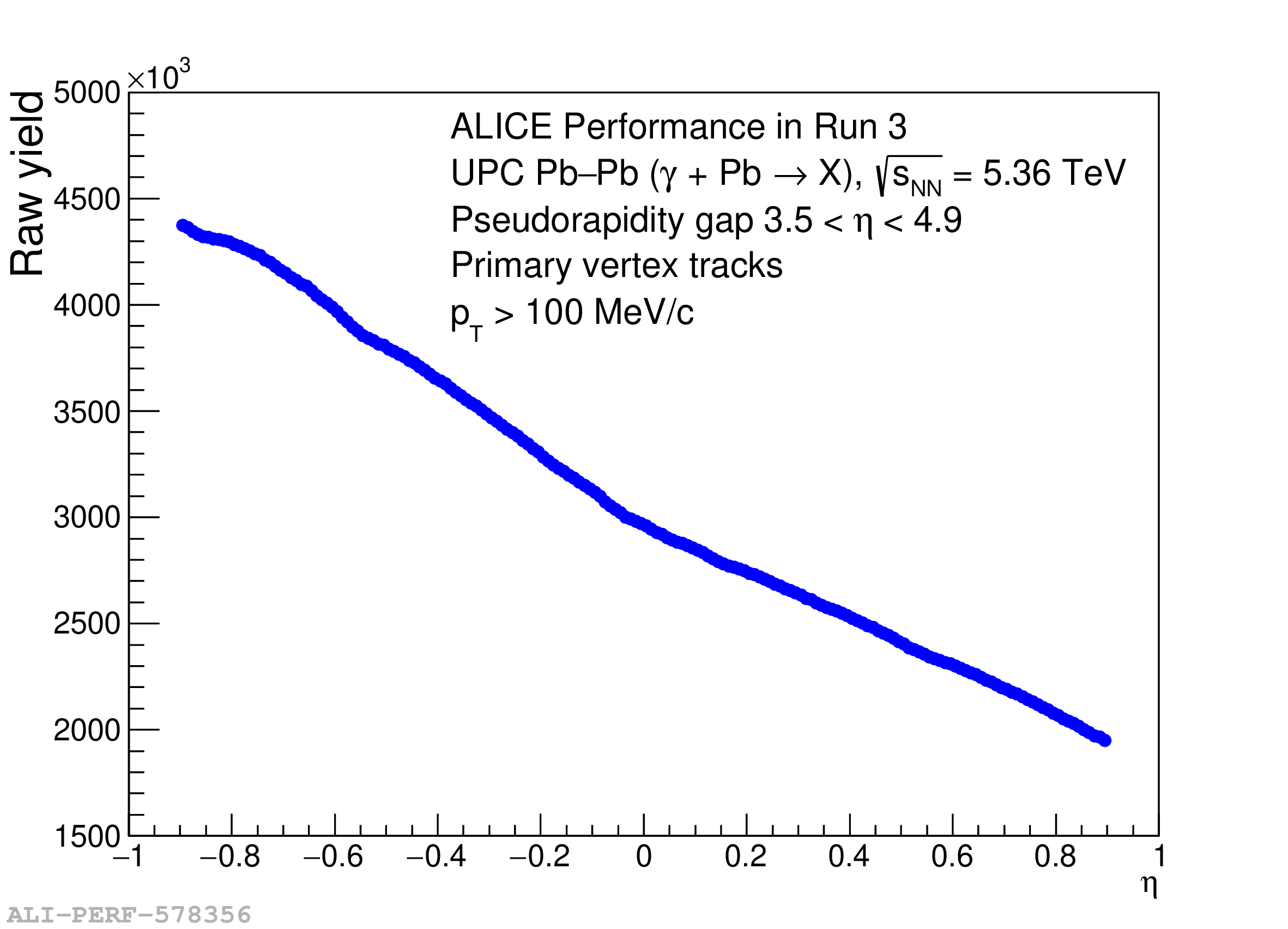}
    \includegraphics[width=0.49\linewidth]{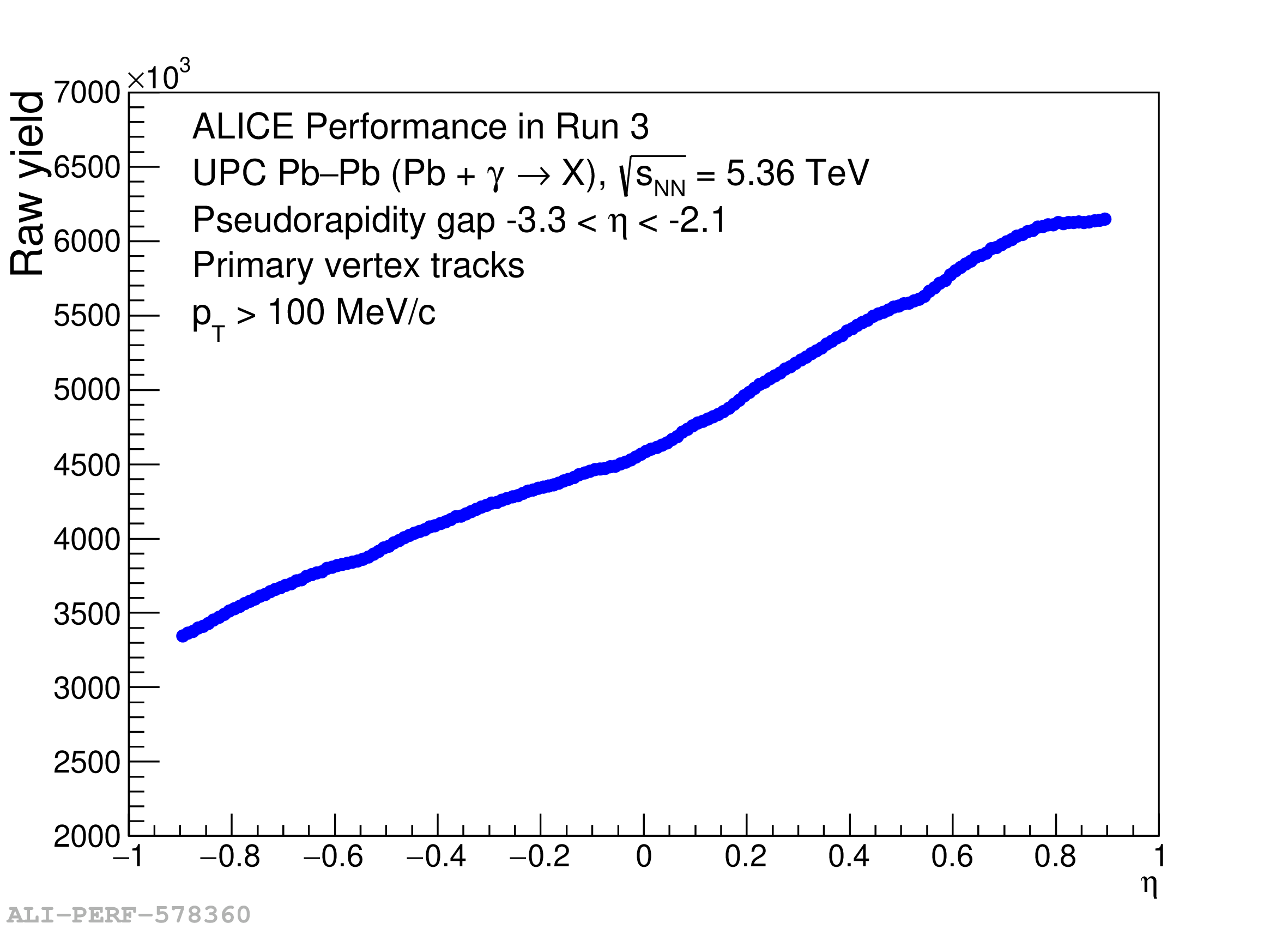}
    \caption{Asymmetric pseudorapidity distributions resulting from asymmetric vetoing of forward detector activity in UPC event selection.}
    \label{fig:eta}
\end{figure}

An example of this enhanced flexibility is shown in Fig.~\ref{fig:eta}, where asymmetric veto conditions lead to a pronounced asymmetry in the pseudorapidity distributions of UPC events. Such selections provide experimental control over the photon–nucleus interaction geometry and allow access to a broader class of photon-induced processes in Run~3. The major breakthrough of Run 3 UPC is the streamed readout and no dedicated exclusive trigger, which allows for the inclusive measurements.

\section{Inclusive Open Charm}

One of the first inclusive measurements of Run~3 UPC program is the inclusive photoproduced open charm. These measurements provide access to the gluon content of nuclei in the perturbative QCD regime and constitute a stringent test of theoretical descriptions of inelastic photonuclear interactions~\cite{opencharm}.

\begin{figure}[h!!]
    \includegraphics[width=0.51\textwidth]{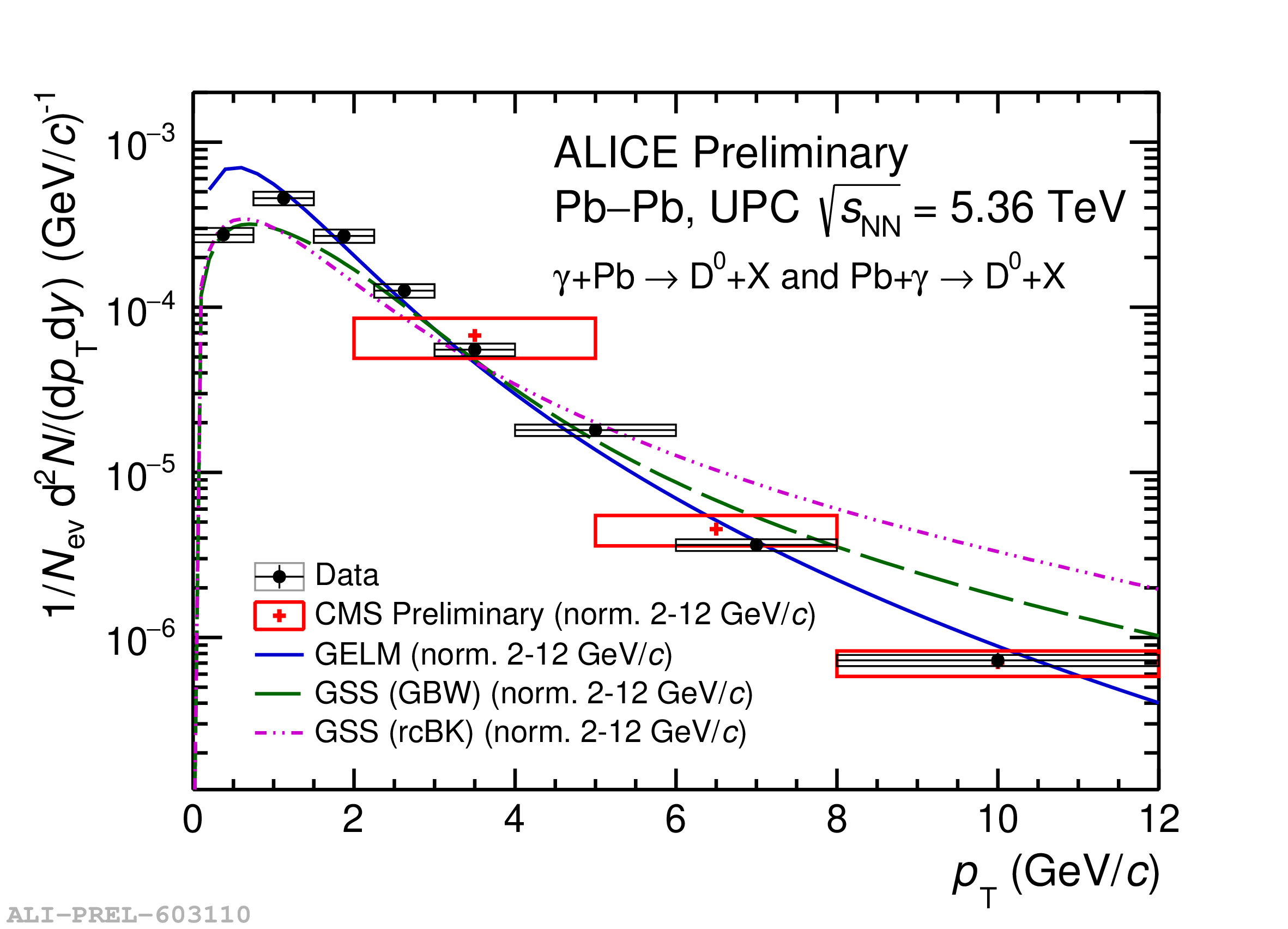}
    \includegraphics[width=0.48\textwidth]{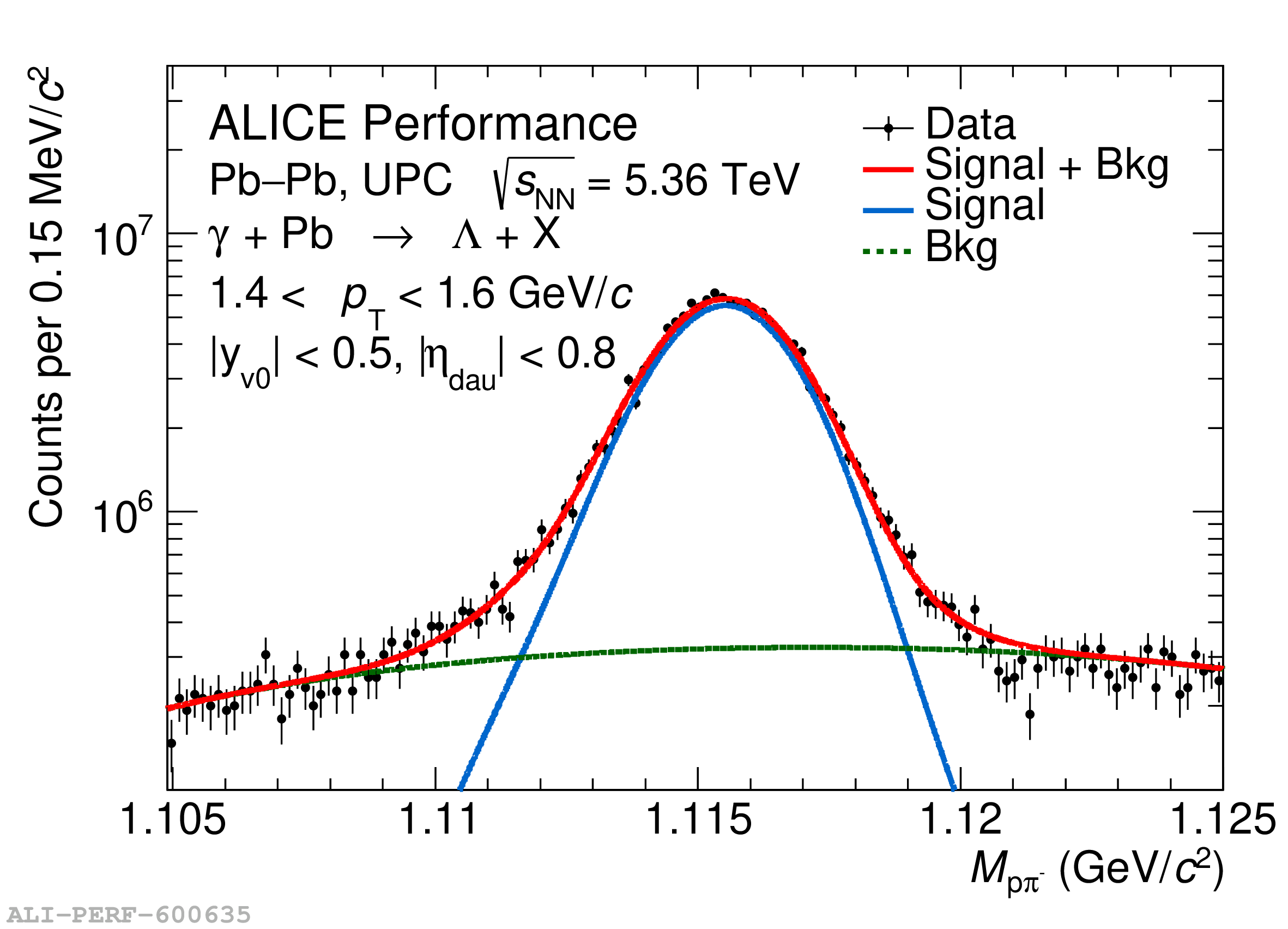}
    \caption{(a) Inclusive D$^{0}$ production as a function of $p_{T}$ in the midrapidity region $|y|< 0.9$, compared with CMS results and theoretical model calculations~\cite{opencharm}. (b) Invariant-mass distribution of reconstructed $\Lambda_{\rm c}^{+}$ candidates.}
    \label{fig:opencharm}
\end{figure}

In Run~3, ALICE has measured inclusive charm production in UPCs down to zero transverse momentum ($p_{\rm T}$), significantly extending the kinematic reach of previous studies. Inelastic UPC events are selected using a single-rapidity-gap topology, enabling the study of photon--nucleus interactions beyond the exclusive regime. Preliminary ALICE results for inclusive open charm, including D-meson production, show significant deviations from predictions based on Color Glass Condensate calculations, indicating that current models do not fully capture the dynamics of inelastic charm photoproduction as illustrated in Fig.~\ref{fig:opencharm} (a)~\cite{opencharm}.

Ongoing analyses aim to improve the precision of these measurements and to extend them to multiple charm hadron species, including D$^{+}$, D$^{*+}$, and $\Lambda_{\rm c}^{+}$ (see Fig.~\ref{fig:opencharm} (b)), as well as complementary channels such as inclusive J/$\psi \rightarrow e^{+}e^{-}$. These studies will provide further constraints on nuclear gluon distributions and help refine theoretical models of photonuclear charm production in UPCs.

\section{Collectivity in Photonuclear Systems}

Inclusive hadron production in UPCs provides a novel opportunity to investigate whether collective effects, traditionally associated with hadronic collisions, can emerge in photon--nucleus interactions~\cite{naturestrangeness,collsmallsystem}. ALICE has performed measurements of identified particle spectra and particle ratios in inclusive $\gamma$--Pb collisions, including pions, kaons, and protons.

\begin{figure}[h!]
    \includegraphics[width=0.52\textwidth]{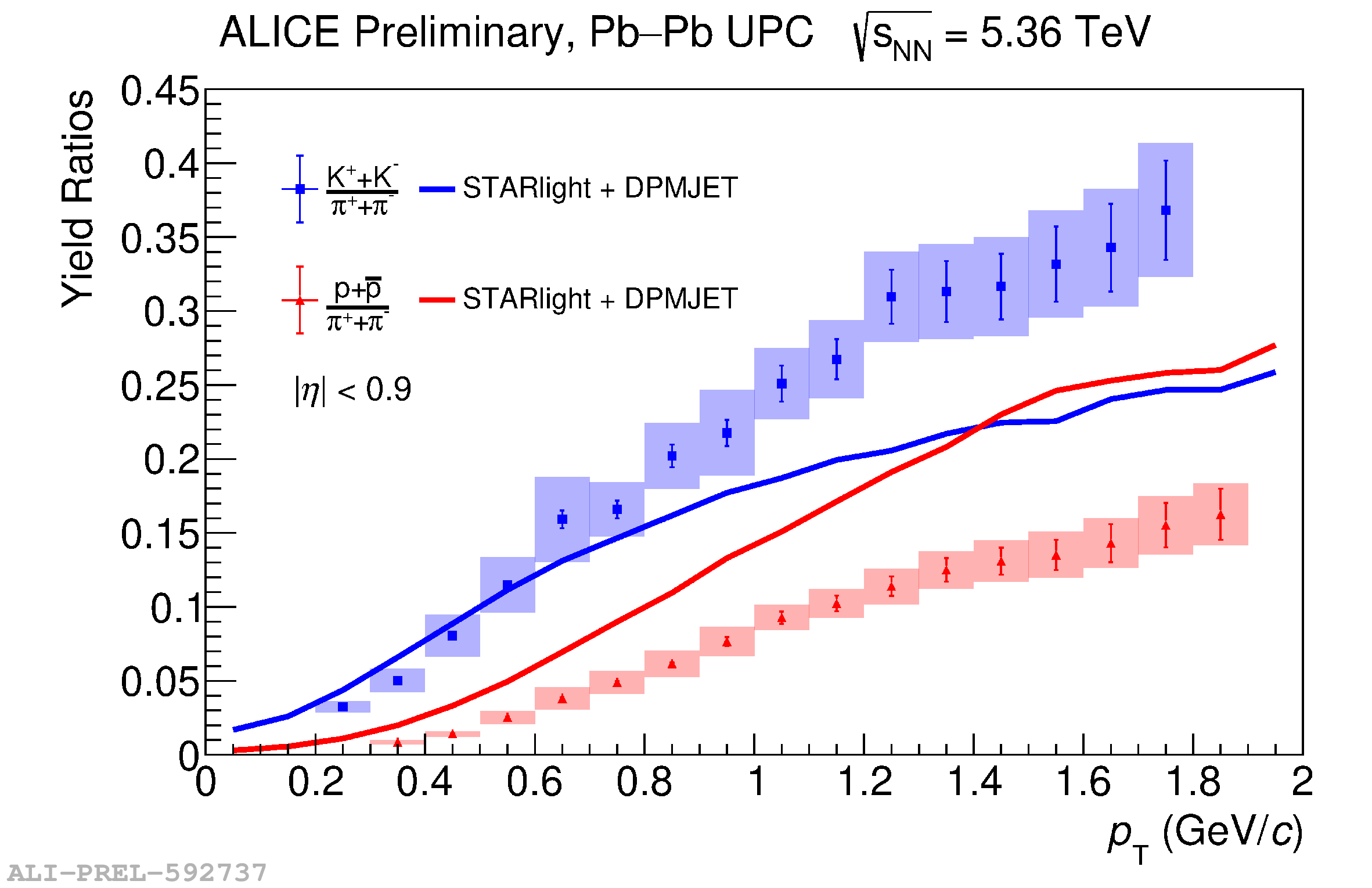}
    \includegraphics[width=0.47\textwidth]{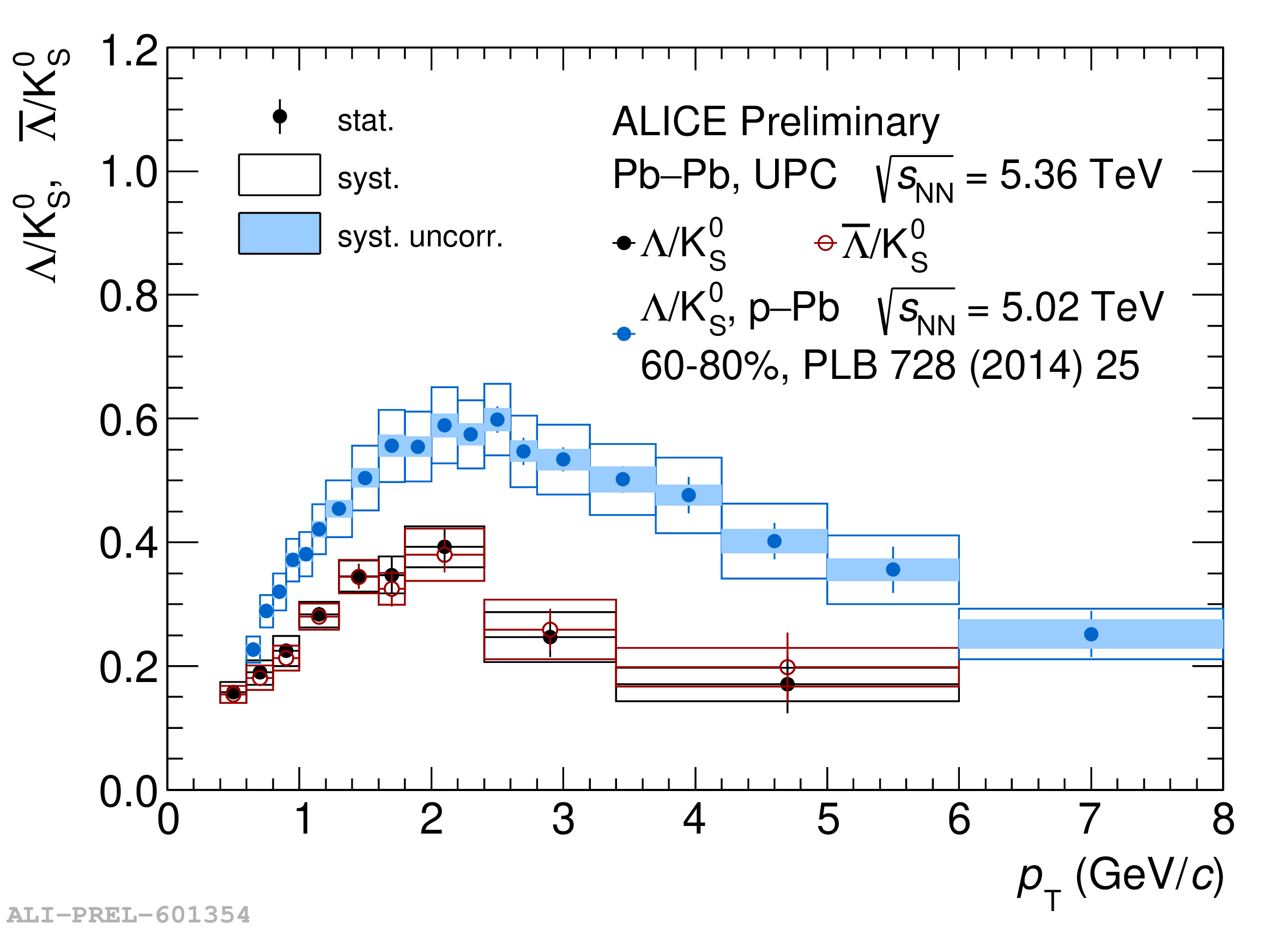}
    \caption{(a) Yield ratios K/$\pi$ and p/$\pi$ in inclusive $\gamma$--Pb UPCs compared to STARlight+DPMJET predictions. 
    (b) Ratio of $\Lambda$ and $\bar{\Lambda}$ to K$^{0}_{\rm S}$ in Pb--Pb UPCs and p--Pb collisions as a function of $p_{\rm T}$.}
    \label{fig:collectivity}
\end{figure}

Preliminary ALICE results for corrected yield spectra show that STARlight+DPMJET describes pion production well, while deviations are observed for heavier hadrons. In particular, the shape of the kaon spectra and the overall normalization of proton yields are not reproduced by the model. Particle ratios such as K/$\pi$ and p/$\pi$, shown in Fig.~\ref{fig:collectivity}(a), further highlight these differences. While the overall trends resemble those observed in pp and p--Pb collisions, STARlight+DPMJET fails to reproduce the measured ratios, indicating missing physics mechanisms in the model description.

The K/$\pi$ ratio in $\gamma$--Pb collisions is found to be slightly lower at low $p_{\rm T}$ compared to hadronic systems, although its $p_{\rm T}$ dependence closely follows that observed in p--Pb and Pb--Pb collisions~\cite{pikpPbPb, pikpPPPbPb}. The p/$\pi$ ratio exhibits a suppression of baryon production with increasing system size, with shapes closer to those observed in pp and p--Pb collisions than in central Pb--Pb interactions. Notably, the low-$p_{\rm T}$ p/$\pi$ ratios approach values measured in central Pb--Pb collisions, hinting at a possible system-size dependence of baryon production mechanisms.

Strangeness production provides further insight into collective-like effects in photonuclear systems. Measurements of $\Lambda$ and K$^{0}_{\rm S}$ production in inclusive $\gamma$--Pb collisions reveal a clear enhancement of the $\Lambda$/K$^{0}_{\rm S}$ ratio around $p_{\rm T} \approx 2~\mathrm{GeV}/c$, as shown in Fig.~\ref{fig:collectivity}(b). The magnitude and shape of this enhancement are similar to those observed in low-multiplicity p--Pb collisions. Moreover, the ratio increases with event multiplicity, evolving from minimum-bias to high-multiplicity $\gamma$--Pb events and approaching values measured in p--Pb collisions. With increasing multiplicity, the $\gamma$--Pb and p--Pb spectra converge, and the overall magnitude of the observed QGP-like signatures is consistent with results reported by ATLAS~\cite{collATLAS}.

Together, these observations suggest that strangeness enhancement and baryon-to-meson modifications, commonly interpreted as signatures of collectivity in hadronic systems, may also manifest in photonuclear interactions. These results challenge the traditional view of UPCs as purely dilute systems and motivate further experimental and theoretical investigations of collective phenomena in small photon-induced systems.

\section{Electroweak Measurements}

Ultra-peripheral collisions provide access to rare electroweak processes through photon--photon interactions at the LHC~\cite{alice3}. In this context, ALICE is exploiting $\gamma\gamma \rightarrow \tau^{+}\tau^{-}$ production to probe the anomalous magnetic moment of the tau lepton, $a_{\tau}=(g-2)/2$. Owing to the large photon flux in heavy-ion UPCs, this process can be studied with competitive sensitivity despite the short lifetime of the $\tau$ lepton.

Several $\tau$-lepton decay topologies are considered in the analysis. These include one-prong decays on both sides (1+1 prong), such as $e+e$ and $e+\mu/\pi$, as well as one- and three-prong decays (1+3 prong), such as $e+3\pi$. The branching fractions are approximately 80\% for one-prong decays and 20\% for three-prong decays. Among these, channels with at least one leptonic decay are favored due to their superior background rejection.

\begin{figure}[h!]
    \includegraphics[width=0.49\textwidth]{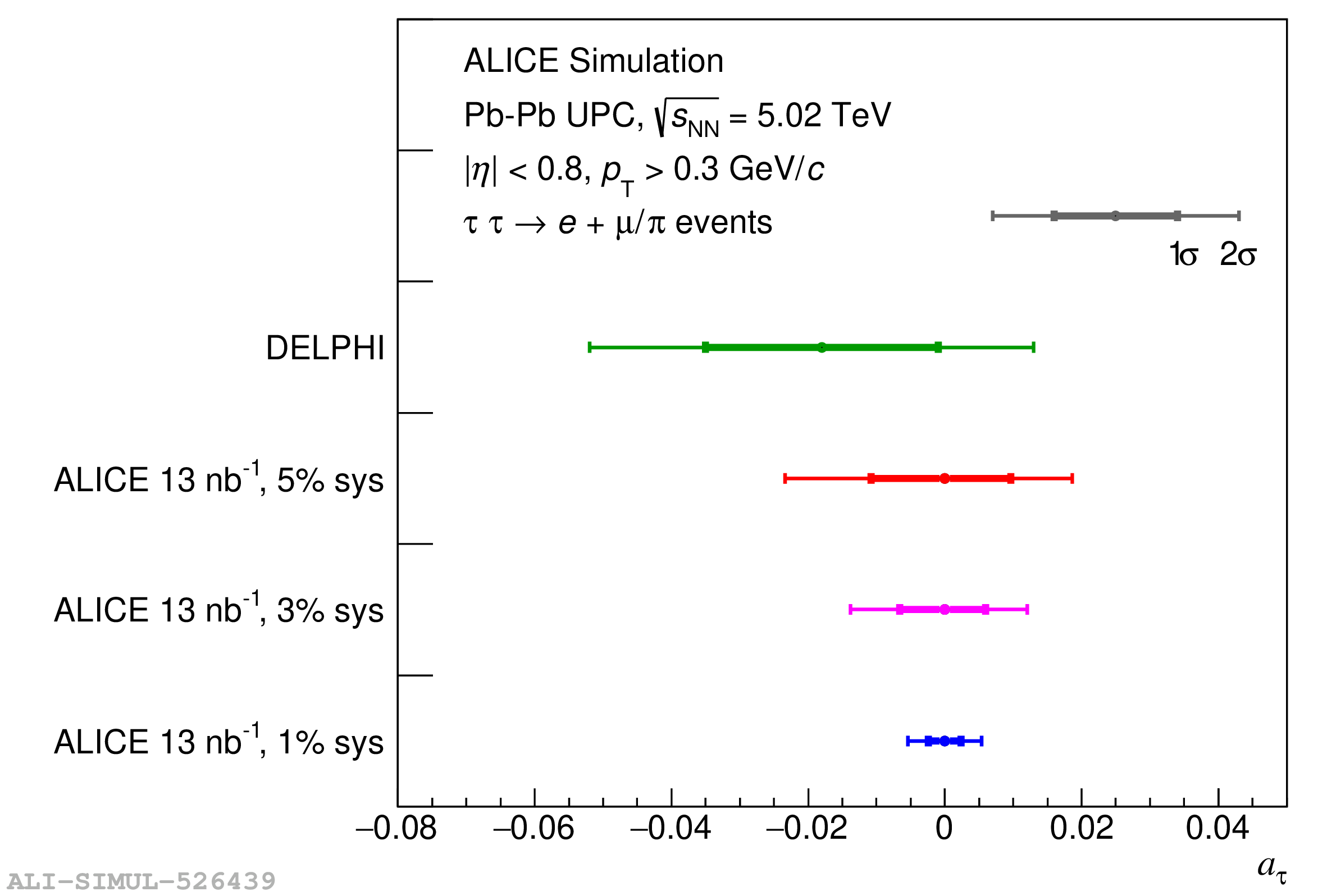}
    \includegraphics[width=0.48\textwidth]{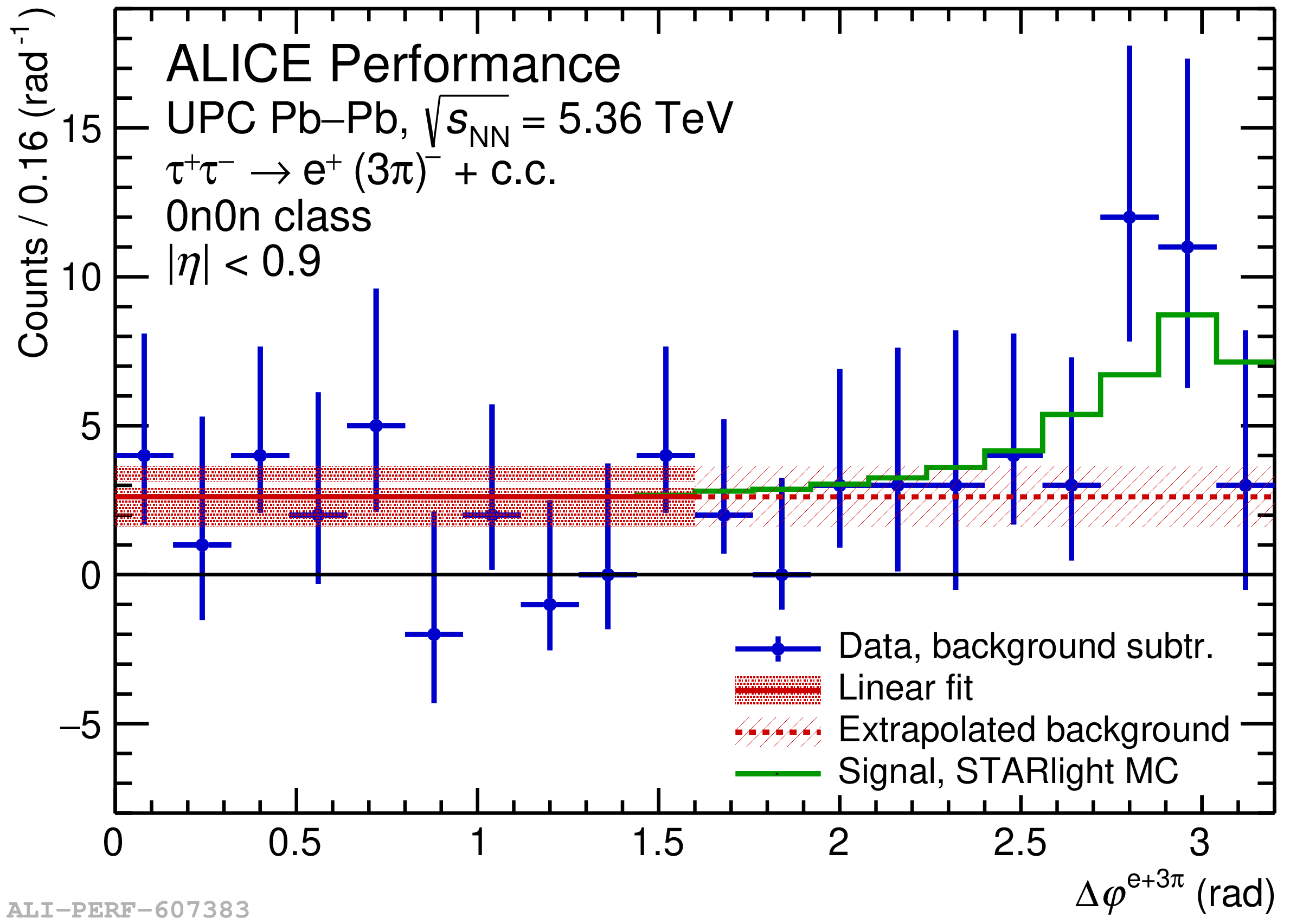}
    \caption{Expected sensitivity to anomalous tau-lepton couplings in $\gamma\gamma \rightarrow \tau^{+}\tau^{-}$ production (left) and representative kinematic distributions used in the event selection (right)~\cite{tautau}.}
    \label{fig:ewk}
\end{figure}

The event selection strategy requires at least one leptonic $\tau$ decay in order to suppress backgrounds from $\gamma\gamma \rightarrow q\bar{q}$ and $\pi\pi$ production. Additional acoplanarity cuts are applied to reduce contributions from continuum dilepton processes. Simulation studies based on Run~3 Pb--Pb conditions with an integrated luminosity of 2.3~nb$^{-1}$ predict approximately $3.6\times10^{4}$ reconstructed events in the central barrel for the $e+\mu/\pi$ topology~\cite{tautau}.

Performance studies indicate that the $e+e$ channel suffers from stronger contamination by back-to-back background processes, while the $e+\mu/\pi$ channel exhibits a fair Monte Carlo description, including non-flat combinatorial backgrounds. The expected yields scale consistently with luminosity and reconstruction efficiencies at the level of a few percent. Overall, the sensitivity to $a_{\tau}$ is driven primarily by the $e+\mu/\pi$ topology, making it the most promising channel for this measurement at ALICE.

Fig.~\ref{fig:ewk} shows the expected sensitivity to anomalous $\tau$-lepton couplings together with representative kinematic distributions used in the event selection, demonstrating the feasibility of this measurement in the UPC environment.

\section{Future Prospects}

The UPC physics program at ALICE will be significantly extended with the installation of the Forward Calorimeter (FoCal) as part of the ALICE upgrade for Run~4, scheduled to start in 2029. FoCal will be located approximately 7~m from the interaction point on the A-side and will provide coverage in the pseudorapidity range $3.4 < \eta < 5.8$ ~\cite{focal}. This new capability will enable measurements of direct photons and exclusive vector meson photoproduction at very forward rapidity, granting access to gluon distributions in nuclei at extremely small Bjorken-$x$.

In particular, FoCal will allow detailed studies of exclusive and dissociative J/$\psi$ and $\psi(2S)$ photoproduction in p--Pb and Pb--Pb UPCs, extending current Run~3 measurements to a previously unexplored kinematic regime. These measurements are expected to provide enhanced sensitivity to gluon saturation effects and to the transverse structure of the nucleus through momentum-transfer-dependent observables. Projections indicate that dissociative J/$\psi$ photoproduction measured in Run~3 can be significantly extended in Run~4 with FoCal acceptance, enabling more differential studies in energy and momentum transfer~\cite{focal, JDdissocuative}.

Looking further ahead, future detector concepts such as ALICE~3, envisaged for Run~5, are expected to significantly enhance the sensitivity of UPC measurements to electroweak processes~\cite{alice3,Alic3upc}. The improved tracking, timing, and extended acceptance foreseen for ALICE~3 would be particularly advantageous for studies of photon--photon interactions, including light-by-light scattering mediated by the box diagram and precision measurements of dilepton final states. These capabilities would enable more stringent tests of Standard Model electroweak predictions and provide sensitivity to possible physics beyond the Standard Model in a complementary kinematic regime.

Together, these developments will open new opportunities for precision studies of photon-induced QCD processes, light-by-light scattering, and electroweak observables, firmly establishing UPCs as a central component of the long-term heavy-ion physics program at the LHC.

\section{Summary}
Ultra-peripheral collisions at ALICE constitute a broad physics program spanning QCD, nuclear structure, and electroweak measurements. Through precision studies of coherent, incoherent, exclusive, and dissociative vector meson photoproduction, ALICE has provided important constraints on gluon shadowing, saturation phenomena, and subnucleonic fluctuations in nuclei. Complementary measurements of spin-related observables, including polarization, azimuthal anisotropies, and quantum interference effects, have demonstrated that UPCs offer access not only to gluon densities but also to the underlying dynamics of vector meson production.

ALICE has also reported novel results on nuclear breakup in UPCs, including isotope production such as gold, mercury, and thallium, highlighting the role of electromagnetic interactions in nuclear transmutation. In addition, inclusive photonuclear measurements have opened new directions, with studies of hadron production suggesting possible collective behavior in photon-induced systems, while open charm photoproduction additionally provides a perturbative QCD probe of gluon dynamics.

Run~3 has marked a significant expansion of the UPC program with the first measurements of inclusive charm and strangeness photoproduction, revealing unexpected features and motivating further theoretical developments. Looking ahead, the installation of FoCal in Run~4 will enable access to an uncharted small-$x$ regime through forward measurements of photon-induced processes, substantially extending the reach of UPC studies. Further in the future, detector concepts such as ALICE~3 will provide enhanced sensitivity to photon--photon interactions, including light-by-light scattering, and precision electroweak measurements. Together, these developments ensure that UPCs will remain a central component of the ALICE heavy-ion physics program in the coming decades.

\end{document}